\documentclass[prb,twocolumn,aps,floatfix,eqsecnum,showpacs]{revtex4}

\usepackage{amssymb}
\usepackage{amsmath}
\usepackage{graphicx}
\usepackage[dvips]{feynmp}
\usepackage{float}
\usepackage[arrow,matrix]{xy}
\usepackage{epstopdf}
\usepackage{hhline}

\begin{document}

\title{Weak localization of bulk channels in topological insulator thin film}

\author{Hai-Zhou Lu and Shun-Qing Shen}
\affiliation{Department of Physics and Centre of Theoretical and Computational Physics, The University of Hong Kong, Pokfulam Road, Hong Kong, China}

\date{\today }

\begin{abstract}
Weak antilocalization (WAL) is expected whenever strong spin-orbit coupling or scattering comes into play. Spin-orbit coupling in the bulk states of a topological insulator is very strong, enough to result in the topological phase transition.
However, the recently observed WAL in topological insulators seems to have an ambiguous origin from the bulk states.
Starting from the effective model for three-dimensional topological insulators, we find that the lowest two-dimensional (2D) bulk subbands of a topological insulator thin film can be described by the modified massive Dirac model. We derive the magnetoconductivity formula for both the 2D bulk subbands and surface bands.
Because with relatively large gap, the 2D bulk subbands may lie in the regimes where the unitary behavior or even weak localization (WL) is also expected, instead of always WAL. As a result, the bulk states may contribute small magnetoconductivity or even compensate the WAL from the surface states. Inflection in magnetoconductivity curves may appear when the bulk WL channels outnumber the surface WAL channels, providing a signature of the weak localization from the bulk states.
\end{abstract}


\maketitle


\section{Introduction}
Topological insulators are materials with gapped bulk states but gapless surface states.\cite{Hasan10rmp,Moore10nat,Qi10rmp}
Due to the topological origin and Dirac fermion nature of the topological surface states,
the topological insulators are expected to have excellent performance in transport.\cite{Roushan09nat,Zhang09prl}
One of intriguing transport features of topological insulators is the weak antilocalization (WAL),
appearing as the negative magnetoconductivity with a sharp cusp in low fields.\cite{Imura09prb,Checkelsky09prl,Peng10NatMat,Checkelsky11prl,Chen10prl,He11prl,Liu11prb,Wang11prb,Liu11arxiv,Chen11rc,Kim11arxiv,Steinberg11arxiv,Lu11arxiv,Tkachov11arxiv,Nestoklon11arxiv}
WAL is intrinsic to topological insulators: (i) so far, most samples have low mobility and long coherence length,
making the quantum interference an important correction to the diffusion transport;
(ii) due to the spin-momentum locking resulted from strong spin-orbit coupling,
the single gapless Dirac cone of the topological surface states carries a $\pi$ Berry phase,\cite{Ando98jpsj,Hsieh09nature}
which changes the interference of time-reversed scattering loops from constructive to destructive.
The destructive interference will give the conductivity an enhancement,
which can be destroyed by applying a magnetic field that breaks the $\pi$ Berry phase, leading to the negative magnetoconductivity with the cusp.

WAL is always expected in systems with either strong spin-orbit scattering or coupling.\cite{HLN80,Bergmann84PhysRep,Iordanskii94jetp,Poole82jpc,Dresselhaus92prl,Chen93prb,Koga02prl,Pedersen99}
Due to vacancies and defects, most as-grown topological insulators have also bulk band carriers which actually dominate the transport.\cite{Hor09prb,Qu10sci}
The bulk bands of topological insulator possess strong spin-orbit coupling, strong enough to invert the normal band structure and give rise to the topological phase transition that defines the nontrivial nature of topological insulators.
One of the simplest choices when considering the conduction bands of a band insulator with spin-orbit coupling that respects time-reversal symmetry is the Rashba model
\begin{eqnarray}\label{rashba}
H_R =\frac{\hbar^2k^2}{2m^*}+\lambda(\sigma_xk_y-\sigma_yk_x),
\end{eqnarray}
where $\hbar$ is the Planck constant over $2\pi$, $(k_x,k_y)$ is the wave vector, $k^2=k_x^2+k_y^2$, $m^*$ is the effective electron mass, and $\sigma_{x,y}$ are Pauli matrices that describe the electron spin. The Rashba model describes two branches of conduction bands, each band has spin locked to momentum, just like the gapless surface states of topological insulator. As a result, the Berry phase for each band of the Rashba model also gives exact $\pi$.\cite{Shen04rc}
According to the $\pi$ Berry phase argument, if the bulk bands of a topological insulator were described by the Rashba model, they should also have the weak antilocalization in the quantum diffusion transport, just like the gapless surface states.

Experimentally, the weak antilocalization is studied by fitting the magnetoconductivity with the Hikami-Larkin-Nagaoka formula,\cite{HLN80}
\begin{equation}\label{HLN}
\Delta \sigma_{\mathrm{HLN}} (B)=\alpha \frac{e^{2}}{\pi h}
\left[ \Psi (\frac{\hbar }{4eB \mathcal{L}^{2}}+\frac{1}{2})-\ln (\frac{\hbar}{4eB\mathcal{L}^{2}})\right] ,
\end{equation}
where $\Psi$ is the digamma function, $B$ is the magnetic field, $\alpha$ and $\mathcal{L}$ are two fitting parameters, $\mathcal{L}$ is an effective phase coherence length.
$\alpha$ is a prefactor, each band that carries a $\pi$ Berry phase should give an $\alpha=-1/2$ prefactor.\cite{Ando98jpsj,McCann06prl}
In the weak interband coupling limit, multiple independent bands with WAL should add up to give bigger negative $\alpha$, e.g., -1, -1.5.
The experimentally fitted $\alpha$ covers a wide range between around -0.4 and -1.1, suggesting that the observed WAL can be interpreted by considering only one or two surface bands,\cite{Checkelsky09prl,Peng10NatMat,Checkelsky11prl,Chen10prl,He11prl,Liu11prb,Wang11prb,Liu11arxiv,Chen11rc,Kim11arxiv,Steinberg11arxiv}
despite of the coexistence of multiple carrier channels from bulk and surface bands at the Fermi surface.
Even, the sharp WAL cusp can be completely suppressed by doping magnetic impurities only on the top surface of a topological insulator.\cite{He11prl}
Why the observed WAL seems to be weakly tied to the bulk bands with strong spin-orbit coupling still poses a mystery to both experimentalists and theorists.
Various interpretations were proposed, such as electron-electron interaction\cite{Wang11prb} and the mixture of the surface states on the top and bottom surfaces of the topological thin film.\cite{Chen11rc}

In this work, we try to investigate the role played by the bulk states in the quantum diffusion transport of a topological insulator thin film.
We find that the two-dimensional (2D) modified Dirac model can provide a unified description for both the surface bands and the lowest 2D bulk subbands of a topological insulator thin film.
We derive the magnetoconductivity formula for the 2D modified Dirac model in the weak interband scattering limit.
In this unified description, whether one has weak antilocalization or weak localization (WL) is governed mainly by the mass (gap) term in the modified Dirac model.
In the massless limit, one has weak antilocalization, while a finite gap can lead to the weak localization or the unitary behavior.
Contrast to the gapless surface states, the bulk states of a topological insulator actually have relatively large gap.
Therefore, while the surface bands probably exhibit weak antilocalization, we suggest that bulk bands may reside in the weak localization or the unitary regime.
The experimentally observed ``weak antilocalization" may be a collective result from both the weak antilocalization of the surface channels and weak localization (or unitary behavior) of the bulk channels. This may help to explain why the fitting parameters of the Hikami-Larkin-Nagaoka formula cover a wide range in the experiments.\cite{Checkelsky09prl,Peng10NatMat,Checkelsky11prl,Chen10prl,He11prl,Liu11prb,Wang11prb,Liu11arxiv,Chen11rc,Kim11arxiv,Steinberg11arxiv}
The unitary behavior (with small magnetoconductivity proportional to the square of the magnetic filed) of bulk channels may help to understand why the bulk states seem ``missing" in the magnetoconductivity although they actually dominate the transport.

The paper is organized as follows. In Sec. \ref{sec:model}, we introduce the 2D modified Dirac model that describes both the surface bands and the lowest 2D bulk subbands of a topological insulator thin film. In Sec. \ref{sec:MC}, the magnetoconductivity formula is presented for the 2D modified Dirac model in the quantum diffusion regime and the weak interband scattering limit. In Sec. \ref{sec:single}, we show the crossover from WAL to WL for a single channel of the modified Dirac model. In Secs. \ref{sec:2bulk} and \ref{sec:morebulk}, we present the total magnetoconductivity of two surface WAL channels and multiple bulk WL channels. Finally, a summary is given in Sec. \ref{sec:summary}.

\section{\label{sec:model}Unified description of bulk and surface of a topological insulator}

\begin{figure}[tbph]
\centering \includegraphics[width=0.5\textwidth]{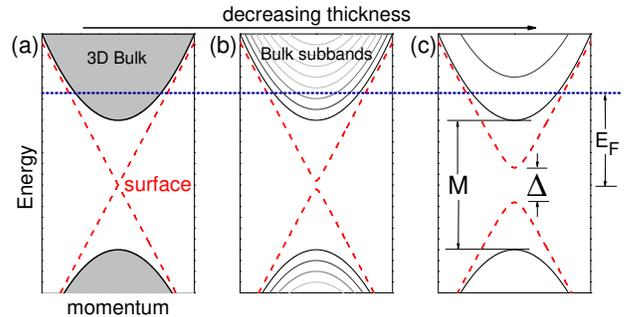}
\caption{(Color online) (a) The gapped bulk (grey area) and gapless surface (dashed lines) bands of a 3D topological insulator.
(b) The quantum confinement along the $z$ direction splits the 3D bulk bands into 2D subbands, while the hybridization of the top and bottom surfaces opens a gap ($\Delta$) for the gapless surface bands. (c) In the ultrathin limit, the Fermi surface intersects with only one pair of bulk subbands (with band gap $M$) and one pair of gapped surface bands (each curve is two-fold degenerate). The horizontal dot line marks the Fermi energy $E_F$ measured from the Dirac point. }
\label{fig:Ek}
\end{figure}

The minimal model to describe a three-dimensional (3D) topological insulator is the modified Dirac model.\cite{Shen11spin}
\begin{eqnarray}\label{H3D}
    H_{\mathrm{3D}}&=&
    \epsilon_{\mathbf{k}}+A\mathbf{k}\cdot \boldsymbol{\alpha}+\mathcal{M}_{\mathbf{k}}\beta,
\end{eqnarray}
where $\mathbf{k}=(k_x,k_y,k_z)$ are wave vectors, the $4\times 4$ Dirac matrices $\boldsymbol{\alpha}=(\alpha_x,\alpha_y,\alpha_z)$, and $\beta$ satisfy the relations
\begin{eqnarray}
\alpha _{i}^{2} =\beta ^{2}=1, \ \alpha _{i}\alpha _{j} =-\alpha _{j}\alpha _{i}, \
\alpha _{i}\beta  =-\beta \alpha _{i}.
\end{eqnarray}
$\epsilon_{{\bf k}}=C+D(k_z^2+k^2)$, $\mathcal{M}_{{\bf k}}=m-B(k_z^2+k^2)$, $k_{\pm}=k_x\pm ik_y$, and $k^2=k_x^2+k_y^2$. $A$, $B$, $C$, $D$, and $m$ are model parameters.\cite{Zhang09np,Liu10prb}
For simplicity, three-dimensional isotropy is assumed. The modified Dirac model constitutes the minimal description of the nontrivial topological nature of band insulators: for $m B >0$, the solution of $d$-dimensional topologically-protected in-gap boundary states (surface states, edge states) can be found by solving the $(d+1)$-dimensional modified Dirac model with open boundary conditions.\cite{Zhou08prl,Lu10prb,Shan10njp,Shen11spin}
Contrast to it, we emphasize that the non-Dirac models with spin-orbit coupling or scattering [\emph{e.g.}, the Rashba model in Eq. (\ref{rashba})] are not enough to describe a topological insulator because they can not give the topologically-protected boundary states solution. Therefore, it is unconsidered to expect the bulk states of topological insulator to have WAL by simply assuming them as non-Dirac electron gases with strong spin-orbit coupling or scattering.

With translational symmetry in all three dimensions, $H_{\mathrm{3D}}$ in Eq. (\ref{H3D}) gives two energy bands [grey area in Fig. \ref{fig:Ek}(a)]
\begin{eqnarray}
E^{\mathrm{3D}}_{\pm\mathbf{k}} &=& \epsilon_{{\bf k}}  \pm \sqrt{\mathcal{M}_{\mathbf{k}}^2 +A^2 (k_z^2+ k^2)},
\end{eqnarray}
separated by the band gap $2m$.
Now we consider the film geometry by imposing the open boundary conditions to the top and bottom surfaces defined along the $z$ direction, topologically-protected surface states will emerge in the gap [red dashed lines in Fig. \ref{fig:Ek}(a)]. For infinite thickness along the $z$ direction (the bulk limit), the surface bands are gapless.
Now if we consider finite thickness along the $z$ direction, the quantum confinement will split the 3D bulk bands $E_{\pm\mathbf{k}}$ into a series of 2D subbands [solid curves in Fig. \ref{fig:Ek}(b)] as $k_z$ is quantized into discrete values.
Meanwhile, a finite-size gap will open for the surface bands [dashed curves in Fig. \ref{fig:Ek}(b)] due to the hybridization of top and bottom surfaces.\cite{Zhou08prl,Linder09prb,Liu10rc,Lu10prb,Shan10njp}
In the ultrathin limit (\emph{e.g.}, three quintuple layers of Bi$_2$Se$_3$ or Bi$_2$Te$_3$ thin films\cite{Zhang10np,Li10am,Liu11prb,Liu11arxiv}),
the Fermi surface may intersect with up to one pair of 2D bulk subbands [see Fig. \ref{fig:Ek}(c)], and the gap $\Delta$ of the surface bands becomes quite visible.\cite{Zhang10np}

\subsection{Topological surface bands}

For the gapped surface bands (dashed curves in Fig. \ref{fig:Ek}), their effective Hamiltonian can be derived from the 3D model in Eq. (\ref{H3D}),\cite{Lu10prb,Shan10njp}
\begin{eqnarray}\label{Hs}
H_{S}=\widetilde{D} k^2+\tau_z(\frac{\Delta}{2}-\widetilde{B}k^2)\sigma_z  + \widetilde{A} (\sigma_xk_y-\sigma_yk_x).
\end{eqnarray}
It has two $2\times2$ blocks, with $\tau_z=\pm 1$ the block index. $\sigma_{x,y,z}$ are the Pauli matrices.
The model parameters $\widetilde{C}$, $\widetilde{D}$, $\Delta$, $\widetilde{B}$, and $\widetilde{A}$ are functions of the thickness.
The gap $\Delta$ for the surface bands is opened at the Dirac point due to the top-bottom surface hybridization.
In general, $\Delta$ increases with decreasing thickness, but could also vanish at some critical thicknesses where there may be topological phase transitions between the quantum spin Hall and normal states.\cite{Linder09prb,Liu10rc,Lu10prb,Shan10njp,Sakamoto10prb}
$\Delta$ and $\widetilde{B}$ become negligible for large enough thickness (e.g., tens of nanometers\cite{Lu10prb}).
$\Delta$ can acquire extra correction in the presence of magnetic doping\cite{Chen10sci,Wray11np} $\Delta\rightarrow \Delta+\tau_z \Delta_F$, where $\Delta_F$ represents a mean-field from the exchange interaction with magnetic impurities. Because block $\tau_z=+1$ and $-1$ have opposite spin definitions, the same magnetic doping will increase the gap in one block while decrease in the other.\cite{Yu10sci}

\subsection{2D bulk subbands}

The simplest way to consider the lowest 2D bulk subbands in Figs. \ref{fig:Ek} (b) and (c) is to replace $\langle k_z\rangle=0$ and $\langle k_z^2\rangle=(\pi/d)^2$, where $d$ is the thickness of the film.
After defining $C+D (\pi/d)^2 =0$, $M/2\equiv m-B (\pi/d)^2$,
the Hamiltonian of the lowest 2D bulk subbands can be written as
\begin{eqnarray}\label{H3Dsub}
    H=D k^2+\tau_z(\frac{M}{2}-Bk^2)\sigma_z  + A (\sigma_xk_x+\sigma_yk_y).
\end{eqnarray}
$\tau_z,\sigma_{x,y,z}$ have the same meanings as in Eq. (\ref{Hs}). Hamiltonians (\ref{Hs}), (\ref{H3Dsub}), and the BHZ model for the quantum spin Hall effect of the HgTe quantum well\cite{Bernevig06science} can be classified as the modified Dirac model in two dimensions.\cite{Shen11spin} Comparing with the original Dirac model, it has an extra $Bk^2\sigma_z$ term, which helps to regulate the boundary properties as $k\rightarrow \infty$ and give well-defined integer Chern number\cite{Lu10prb} and $Z_2$ index.\cite{Shan10njp}
A set of parameters for this model is given in Table \ref{tab:thinfilm parameter} for 5 nm.
\begin{table}[htb]
  \centering
  \begin{minipage}[t]{\linewidth}
      \caption{The parameters in Eq. (\ref{H3Dsub}) for $d=5$ nm, calculated from the model parameters of the effective model for 3D topological insulators.\cite{Liu10prb} $k_C\equiv \sqrt{M/2B}$.}
\label{tab:thinfilm parameter}
\begin{tabular}{lccc}
    \hline\hline
    & Bi$_2$Se$_3$ & Bi$_2$Te$_3$ & Sb$_2$Te$_3$ \\\hline
       $D$ (eV$\cdot$\AA$^2)$ & 30.4 & 49.68 & -10.78 \\\hline
       $M$  (eV) & -0.5 & -0.58 & -0.28 \\\hline
       $B$ (eV$\cdot$\AA$^2)$ & -44.5 & -57.38 & -48.51 \\\hline
       $A$ (eV$\cdot$\AA) & 3.33 & 2.87 & 3.40 \\\hline
       $k_C$ (\AA$^{-1})$ & 0.075 & 0.071 & 0.054 \\\hline
       \hline
  \end{tabular}
\end{minipage}
\end{table}

In the absence of the $Dk^2$ and $Bk^2\sigma_z$ terms and for only one block of the modified Dirac model, the quantum diffusion transport is known.\cite{Imura09prb,Lu11arxiv}
In the gapless limit ($M=0$), there is only weak antilocalization with negative magnetoconductivity cusp.\cite{Suzuura02prl,McCann06prl} A finite
gap $M$ will lead to a crossover from weak antilocalization to: (i) the unitary regime with small magnetoconductivity proportional to the square of magnetic field for moderate gap/Fermi energy ratios; and to (ii) weak localization with positive magnetoconductivity cusp for large gap/Fermi energy ratios.\cite{Imura09prb,Lu11arxiv}

In contrast to the surface bands with no or small gaps, the 2D bulk subbands of a 3D topological insulator have relatively large band gaps (more than hundreds of meV, see Table. \ref{tab:thinfilm parameter}).
As a result, instead of always weak antilocalization, the unitary behavior or weak localization is likely expected for the bulk states of a topological insulators.
Moreover, $Dk^2$ and $Bk^2\sigma_z$ will bring more scenarios.

\section{\label{sec:MC}Magnetoconductivity for 2D modified Dirac model}

We have shown that the 2D modified Dirac model in Eq. (\ref{H3Dsub}) provides a unified description for (i) the 2D bulk subbands due to the quantum confinement, and (ii) the surface bands of a 3D topological insulator.
The quantum diffusion transport of the bulk subbands and surface bands can be studied starting from this model.
It has two $2\times2$ blocks with block index $\tau_z=\pm$, each block has a conduction band and a valence band, denoted as $|\tau_z, c/v\rangle$, with the band dispersions given by
$E_{c/v}=Dk^2\pm\sqrt{(M/2-Bk^2)^2+A^2k^2}$, where $c$ and $v$ stand for conduction and valence bands, respectively.
Without loss of generality, we assume that the transport is only contributed by the conduction bands that cross the Fermi surface.
Each band contributes one \emph{channel}. If we assume that the interband coupling by impurity scattering is weak, the total conductivity is the summation of that from each band intersecting the Fermi surface.
The conduction bands $|+,c\rangle$ and $|-,c\rangle$ should have the same magnetoconductivity formula. The wavefunction of $|+,c\rangle$ is given by
$|+,c\rangle=(
    \cos\frac{\Theta}{2},\sin\frac{\Theta}{2} e^{i\varphi}
)^T e^{i\mathbf{k}\cdot\mathbf{r}}/\sqrt{S}$, where $\tan\varphi \equiv k_y/k_x$, and at the Fermi surface
\begin{eqnarray}\label{costheta}
\cos\Theta \equiv \frac{M/2-Bk_F^2}{ E_F-Dk_F^2 },
\end{eqnarray}
where $E_F$ and $k_F$ are the Fermi energy and Fermi wave vector, respectively.

The weak localization and weak antilocalization happen in the quantum diffusion regime, where the mean free path ($\ell_e$) due to elastic scattering is much shorter than the sample size, but the phase coherence length ($\ell_{\phi}$) due to inelastic scattering is comparable with the sample size.
The quantum diffusion transport can be studied by the standard diagrammatic technique.\cite{HLN80,Altshuler80prb,Bergmann84PhysRep,Suzuura02prl,McCann06prl}
WL and WAL are most evident in the magnetoconductivity, which is defined as the conductivity change as a function of an applied magnetic field $B$.
In this work, we consider that the conductivity is measured along the $x$ direction and the magnetic field is applied only along $z$ direction.
In the calculation, we consider the nonmagnetic elastic scattering by static centers as well as random scattering by magnetic impurities.
The calculation is similar to that for the topological surface states with a magnetically-doped gap,\cite{Lu11arxiv} where $\cos\Theta\equiv \Delta/2E_F$ compared to Eq. (\ref{costheta}), so we present only the result here.
For band $|+,c\rangle $, the zero-temperature magnetoconductivity is found as
\begin{equation}\label{MC}
\Delta \sigma (B)=\sum_{i=0,1}\frac{\alpha _{i}e^{2}}{\pi h}
\left[ \Psi (\frac{\ell_{B}^{2}}{\ell _{\phi i}^{2}}+\frac{1}{2})-\ln (\frac{\ell _{B}^{2}}{\ell_{\phi i }^{2}})\right] ,
\end{equation}
where $\Psi $ is the digamma function, $\ell _{B}^{2}\equiv \hbar/(4e|B|) $ is the magnetic length, $1/\ell_{\phi i}^2\equiv 1/\ell_{\phi}^2+1/\ell^2_i$, $\ell_{\phi}$ is the phase coherence length when $\cos\Theta=0$,
\begin{eqnarray}\label{aLg}
&&\alpha_{0}=\frac{\eta _{v}^{2}(1+2\eta _{H})}{2(1+1/g_{1})},\ \alpha _{1}=-\frac{\eta _{v}^{2}(1+2\eta _{H})}{2(1+1/g_{0}+1/g_{2})},  \notag \\
&&
\ell _{0}^{-2}=\frac{g_{0}}{2\ell^2  (1+1/g_{1})}\ \ell _{1}^{-2}=\frac{g_{1}}{2\ell^2
(1+1/g_{0}+1/g_{2})},\nonumber\\
&&g_{0} \equiv 2[\frac{a^{4}+b^{4}}{a^{4}}\frac{1/\ell^2 }{(1/\ell_e^2+1/\ell_z^2)}-1],  \notag \\
&&g_{1} \equiv 2[\frac{1/\ell^2 }{(1/\ell_e^2-1/\ell_z^2)(2a^{2}b^{2})/(a^{4}+b^{4})-2/\ell^2_x}-1],  \notag \\
&&g_{2} \equiv 2[\frac{a^{4}+b^{4}}{b^{4}}\frac{1/\ell^2 }{(1/\ell_e^2+1/\ell_z^2)}-1],
\end{eqnarray}
$a\equiv \cos\frac{\Theta}{2}$, $b\equiv\sin\frac{\Theta}{2}$. Also,
$1/\ell^2\equiv1/\ell_e^2+1/\ell_m^2$, where $\ell_e$ is the mean free path. The calculation assumes short mean free path ($\ell
_{e}\ll \ell_B$).
$\ell_m$ is the magnetic scattering length that characterizes the strength of the random scattering
by magnetic impurities, $1/\ell_m^2=1/\ell_x^2+1/\ell_y^2+1/\ell_z^2$, $\ell_{x,y,z}$ are components along the $x,y,z$ directions, respectively. In-plane isotropy ($\ell_x=\ell_y$) is assumed in Eq. (\ref{MC}).
Shorter $\ell_m$ means stronger magnetic scattering. In Eq. (\ref{MC}), $\eta _{v}=\left[ 1-a^{2}b^{2}(\ell^2 /\ell_e^2-\ell^2 /\ell_z^2)
/(a^{4}+b^{4})\right] ^{-1}$ comes from the correction to velocity from the ladder diagrams,\cite{Shon98jpsj}
and $\eta_{H}=-( 1-1/\eta _{v}-\ell^2 /\ell_x^2)/2$ comes from the dressed Hikami boxes.\cite{McCann06prl}
The lengthes are related to characteristic times by $\ell_e^2=\mathcal{D}\tau_e$ and $\ell_m^2=\mathcal{D}\tau_m$, where $\tau_e$ is the elastic scattering time, $\tau_m$ is the magnetic scattering time. The defined diffusion coefficient $\mathcal{D}=v_F^2\tau/2$, where $1/\tau\equiv 1/\tau_e+1/\tau_m$ and the Fermi velocity $v_F$ is given by
\begin{eqnarray}
v_{\mathrm{F}}=\frac{1}{\hbar}(2Dk_F+ A\sin\Theta -2Bk_F \cos\Theta ).
\end{eqnarray}

\section{Results}

\subsection{\label{sec:single}Single channel}

\begin{figure}[tbph]
\centering \includegraphics[width=0.5\textwidth]{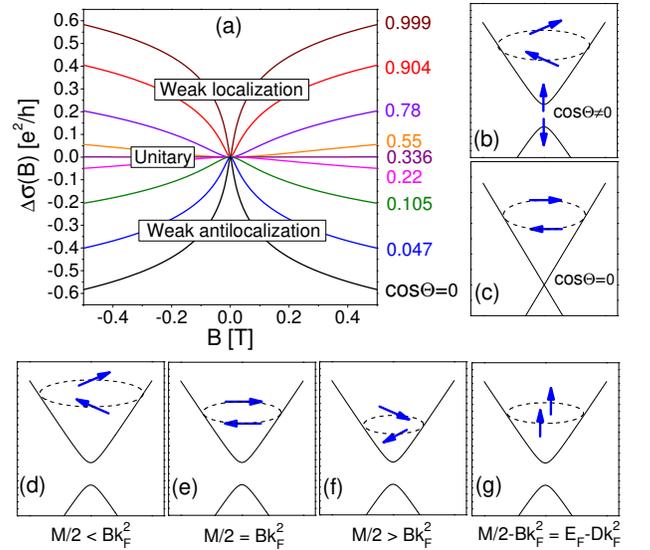}
\caption{(Color online) (a) Magnetoconductivity of a single channel of the 2D modified Dirac model. A crossover from weak antilocalization to weak localization is expected when increasing $\cos\Theta$. $\ell_{\phi}=300$ nm and $\ell_e=10$ nm. Weak isotropic magnetic scattering is assumed, $\ell_x/\sqrt{3}=\ell_z/\sqrt{3}=\ell_m=10000\sqrt{3}$ nm.
(b)-(g) Possible pseudospin patterns of a conduction band of the modified Dirac model.}
\label{fig:mc_Dirac}
\end{figure}

We have shown that for each channel contributed by the modified Dirac model, the magnetoconductivity is given by Eq. (\ref{MC}).
Despite of many parameters in Eq. (\ref{aLg}), the most decisive one is $\cos\Theta$ defined in Eq. (\ref{costheta}).
As shown in Fig. \ref{fig:mc_Dirac}, the magnetoconductivity formula describes a crossover from weak antilocalization to weak localization, controlled by $\cos\Theta$.
In the limit $\cos\Theta=0$, one has only WAL with negative magnetoconductivity cusp.
Increasing $\cos\Theta$ will drive the system first into the unitary regime with small magnetoconductivity proportional to $B^2$, and finally to WL with positive magnetoconductivity cusp when $\cos\Theta\rightarrow 1$. In the limit $\cos\Theta =0$, one has $\alpha_0=0$ and $\alpha_1=-1/2$;
In the limit $\cos\Theta\rightarrow 1$, one has $\alpha_0=1/2$ and $\alpha_1=0$, corresponding to that the prefactor $\alpha$ in the Hikami-Larkin-Nagaoka formula goes from $-1/2$ to $1/2$.

The $\pi$ Berry phase explained the weak antilocalization of a single gapless Dirac cone,\cite{Ando98jpsj}
the crossover as the function of $\cos\Theta$ can be understood similarly.
For the conduction band $|+,c\rangle$, the Berry phase at the Fermi surface is found as
\begin{eqnarray}
-i\int_0^{2\pi} d\varphi \langle +,c| \frac{\partial}{\partial \varphi} |+,c\rangle  =\pi (1-\cos\Theta).
\end{eqnarray}
where $\cos\Theta$ is given by Eq. (\ref{costheta}).
Therefore, one has the $\pi $ Berry phase for WAL when $\cos\Theta=0$ and $0$ Berry phase for WL when $\cos\Theta\rightarrow 1$.
Different from the Berry phase in absence of $Bk^2\sigma_z$ term,\cite{Ghaemi10prl} one may also have the $\pi $ Berry phase as well as the weak antilocalization at $2\pi \rho = k_F^2= M/2B$, where $\rho$ is the sheet carrier density of investigated band.\cite{Tkachov11arxiv} Moreover, $Dk_F^2$ may effectively reduce $E_F$ to $E_F-Dk_F^2$, reminding that the effective model for the gapless surface states also contains the $Dk^2$ term,\cite{Shan10njp,Culcer10prb} this may explain the earlier crossover to weak localization at relatively large $E_F$.\cite{Liu11arxiv}
Because the Berry phase depends on $\cos\Theta$, the crossover can also be understood by the pseudospin orientation along $z$ direction.
Several typical orientation cases are given in Figs. \ref{fig:mc_Dirac}(b)-(g).
WAL with $\alpha=-1/2$ corresponds to purely in-plane polarization, as in Figs. \ref{fig:mc_Dirac}(c) and (e);
WL with $\alpha=1/2$ corresponds to the fully polarized along $z$, as in Figs. \ref{fig:mc_Dirac} (b) and (g).
Due to the $Bk^2_F\sigma_z$ term, one may expect purely in-plane polarization at $k_F=\sqrt{M/2B}$, and the pseudospin polarization can change from $z$ to $-z$ direction as shown in Figs. \ref{fig:mc_Dirac}(d)-(f). Due to the $Dk_F^2$, one may expect the full polarization along $z$ at finite $k_F$ as shown in Fig. \ref{fig:mc_Dirac}(g), instead of always at the bottom of the band (Fig. \ref{fig:mc_Dirac}b).

The Berry phase of the valence bands is given by $\pi(1+\cos\Theta)$.
Therefore, if the sample is $p$-type, i.e., the Fermi surface is at the 2D valence subbands,
the quantum diffusion transport will give the similar competition between WL and WAL, as long as the 2D valence subbands are also well described by the massive Dirac model (situations in Bi$_2$Se$_3$ and Bi$_2$Te$_3$ may be more complicated because of their valence band maxima away from the $\Gamma$ point).

Recently the quantum diffusion transport was also theoretically studied for the HgTe quantum well,\cite{Tkachov11arxiv}
where the suppression of weak antilocalization (corresponding to the unitary behavior in this work) was also found for the BHZ model.
Beyond the suppression of the weak antilocalization, we also expect the weak localization in the HgTe quantum well because the BHZ model can also be classified as the modified Dirac model.

Because the two blocks of the modified Dirac model in Eq. (\ref{H3Dsub}) are mutually time-reversal partners,
the crossover here does not break time-reversal symmetry, different from those in the ferromagnetic semiconductors with spin-orbit coupling\cite{Dugaev01prb,Neumaier07prl} and
magnetically-doped topological surface states.\cite{Liu11arxiv,Lu11arxiv}

\subsection{\label{sec:2bulk}Two surface channels and two bulk channels}

\begin{figure}[tbph]
\centering \includegraphics[width=0.49\textwidth]{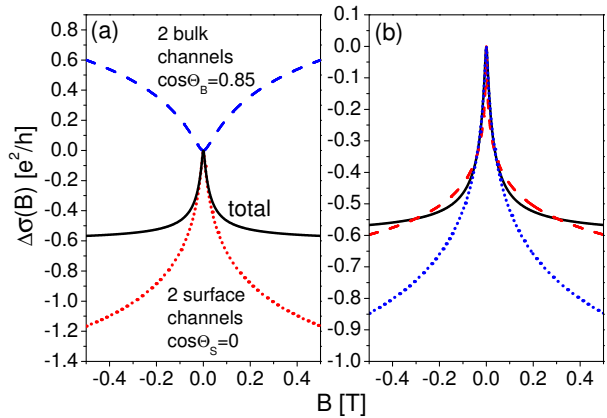}
\caption{(a) The magnetoconductivity of two surface channels with $\cos\Theta_S=0$ (dotted line), two bulk channels with $\cos\Theta_B=0.85$ (dashed line), and their summation (solid line). (b) solid: the same as the solid line in (a). Dash line: fitting to solid line between $B\in[-0.5, 0.5]$ with $\alpha=-0.316$ and $\mathcal{L}=941$. Dotted line: fitting to solid line between $B\in[-0.05, 0.05]$ with $\alpha=-0.659$ and $\mathcal{L}=363$. All the parameters are given in Fig. \ref{fig:mc_Dirac}.}
\label{fig:fitting}
\end{figure}

\begin{figure}[tbph]
\centering \includegraphics[width=0.5\textwidth]{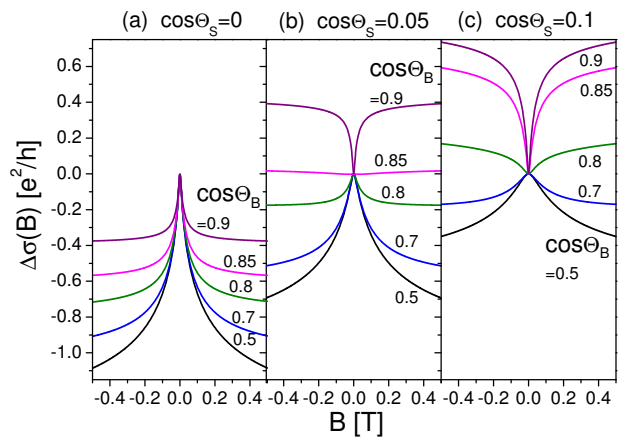}
\caption{Total magnetoconductivity of two surface channels in the weak antilocalization regime and two bulk channels in the unitary or weak localization regime. All the parameters are given in Fig. \ref{fig:mc_Dirac}.}
\label{fig:mc_twobulks}
\end{figure}

\begin{table}[htb]
  \centering
  \begin{minipage}[t]{\linewidth}
      \caption{The Fermi energies of $n$-type as-grown bulk crystals and thin films of topological insulators from the ARPES measurements.
      The Fermi energies are measured from the bottom of the conduction band. }
\label{tab:ARPES}
\begin{tabular}{cccc}
    \hline\hline
   Ref. & sample & thickness & Fermi energies (eV)\\\hline
 \cite{Xia09np,Hsieh09nature,Wray11np}  &   Bi$_2$Se$_3$ & bulk & 0.1   \\\hline
 \cite{Hor09prb}  &  Bi$_2$Se$_3$ & bulk & 0.05   \\\hline
 \cite{Chen09science} &     Bi$_2$Te$_3$ & bulk & 0.045   \\\hline
 \cite{Chen10sci} &     Bi$_2$Se$_3$ & bulk & 0.15 \\\hline
 \cite{Zhang10np} &  Bi$_2$Se$_3$    & 2$\sim$5 QL &  0.1$\sim$0.2   \\\hline
 \cite{Li10am} &     Bi$_2$Te$_3$ & 80 nm & 0.03   \\
  &      & 2-5 QL &  0.1$\sim$0.2   \\\hline
 \cite{Liu11arxiv} &     Bi$_2$Se$_3$ & 3 QL & 0.11  \\\hline
       \hline
  \end{tabular}
\end{minipage}
\end{table}

\begin{table}[htb]
  \centering
  \begin{minipage}[t]{\linewidth}
      \caption{Fitting the weak localization curves in Fig. \ref{fig:mc_twobulks} by the Hikami-Larkin-Nagaoka formula in Eq. (\ref{HLN}). $\alpha$ and $\mathcal{L}$ are two fitting parameters. $B$ are the perpendicular magnetic field in units of Tesla. }
\label{tab:fitting}
\begin{tabular}{cc|cc|ccc}
    \hline\hline
  & & $B\in$ & [-0.5, 0.5] & $B\in$ & [-0.05, 0.05] \\ \hline
 $\cos\Theta_S$ & $\cos\Theta_B$ & $\alpha$ & $\mathcal{L}$ (nm) & $\alpha$ & $\mathcal{L}$ (nm) \\ \hline
 0 & 0.5 & -0.883 & 341 & -0.991 & 301\\ \hline
 0 & 0.7 & -0.650 & 468 & -0.944 & 309\\ \hline
 0 & 0.8 & -0.444 & 683 & -0.815 & 331\\ \hline
 0 & 0.85 & -0.316 & 941 & -0.659 & 363\\ \hline
 0 & 0.9 & -0.184 & 1502& -0.419 & 433\\ \hline
 0.05 & 0.5 & -0.787 & 193 &  -0.945 & 169 \\ \hline
 0.05 & 0.7 & -0.489 & 266 &  -0.830 & 175 \\ \hline
 0.05 & 0.7 & -0.119 & 582 & -0.315  & 221 \\ \hline
 0.1 & 0.5 & -0.594 & 117 & -0.829  & 98.5 \\ \hline
 0.1 & 0.7 & -0.211 & 181 & -0.505  & 108 \\ \hline
       \hline
  \end{tabular}
 \end{minipage}
\end{table}

Now we consider the case shown in Fig. \ref{fig:Ek}(c), with two channels of surface bands and two channels of bulk subbands (note that each curve in Fig. \ref{fig:Ek} represents two degenerate bands labeled by the block index $\tau_z$).
We have shown that for each channel of the modified Dirac model, the magnetoconductivity is given by Eq. (\ref{MC}), which describes a crossover from weak antilocalization to weak localization with increasing $\cos\Theta$.
In the weak interband scattering limit, the total magnetoconductivity is the summation of that from each band.
Similar to Eq. (\ref{costheta}), we introduce two control parameters
\begin{eqnarray}\label{costheta_S}
\cos\Theta_B = \frac{M/2-Bk_F^2}{ E_F-Dk_F^2 },\nonumber\\
\cos\Theta_S = \frac{\Delta/2-\widetilde{B}k_F^2}{ \widetilde{E}_F-\widetilde{D}k_F^2 },
\end{eqnarray}
to characterize the bulk subbands and the surface bands, respectively. The parameters in Eq. (\ref{costheta_S}) were defined in Hamiltonians (\ref{Hs}) and (\ref{H3Dsub}).
Note that the Fermi energies $E_F$ and $\widetilde{E}_F$ are measured from the Dirac points, which may be defined differently for the bulk and surface bands.
In absence of magnetic impurities, two surface channels have the same $\cos\Theta_S$, and two bulk channels have the same $\cos\Theta_B$.
Except in the ultra thin limit (usually less than 5 nm in thickness), the surface states has ignorable $\Delta$ and $\widetilde{B}$.\cite{Lu10prb,Zhang10np,Li10am,Liu11prb}
It is fair to expect small $\cos\Theta_S$ that gives weak antilocalization.
On the other hand, we expect relatively large $\cos\theta_B$ that gives the unitary behavior or weak localization.
When the Fermi surface is near the bottom of the band, the gap $M$ and the Fermi energy $E_F$ dominate $\cos\Theta_B$.
The bulk subbands have relatively large gap, about 0.5 eV for Bi$_2$Se$_3$ and Bi$_2$Te$_3$ (note that this gap is measured at the $\Gamma$ point,\cite{Liu10prb} not the smaller indirect gaps which are irrelevant to the topological nature\cite{Xia09np,Zhang09np,Chen09science}).
The Fermi surfaces of as-grown samples are usually less than 0.20 eV measured from the bottom of the bulk conduction band
(see Table \ref{tab:ARPES}), i.e., $E_F$ is about 0.25 to 0.5 eV measured from the Dirac point.
According to these data, $\cos\Theta_B$ ranges between 0.5 and 1, \emph{i.e.}, between the unitary and weak localization regimes according to Fig. \ref{fig:mc_Dirac}.

In Fig. \ref{fig:fitting}(a), we show the total magnetoconductivity for two surface and two bulk channels, with $\cos\Theta_S=0$ and $\cos\Theta_B=0.85$.
Although the surface and bulk channels exhibit the weak antilocalization and the weak localization, respectively, they collectively behave like the weak antilocalization. Therefore, in Fig. \ref{fig:fitting}(b), we try to fit the total magnetoconductivity by the Hikami-Larkin-Nagaoka formula.
Although widely exploited in fitting experiments,
this formula may not be suitable for multiple channels, because the magnetoconductivity consists of two competing terms even for single channel of the modified Dirac model, not to mention multiple channels.
Therefore, no ubiquitous fitting result of $\alpha$ and $\mathcal{L}$ can be reached when changing the fitting range.
For $B\in[-0.5, 0.5]$, we obtain $\alpha=-0.316$ and $\mathcal{L}=941$, and for $B\in[-0.05, 0.05]$, $\alpha=-0.659$ and $\mathcal{L}=363$.
Both fittings can not reproduce the magnetoconductivity well in the whole range.

In Fig. \ref{fig:mc_twobulks}, we show more general cases with $\cos\Theta_S\in\{0, 0.05, 0.1\}$ and $\cos\Theta_B\in\{0.5, 0.7, 0.8, 0.85, 0.9 \}$.
They are fitted by the Hikami-Larkin-Nagaoka formula in Eq. (\ref{HLN}).
Note that the fitting results almost cover the experimentally recorded $\alpha$ ($\in[-1.1, -0.38]$, see Table \ref{tab:WAL}).
These fitting results imply that one can approach to $\alpha\sim -1/2$ anyway, because of the freedom in choosing $\mathcal{L}$ and fitting range.
Therefore, $\alpha\sim -1/2$ does not always mean that there is only one surface channel of single massless Dirac cone.
It may happen to be the collective result of multiple surface and bulk channels.
In Figs. \ref{fig:mc_twobulks}(b) and (c), one even has weak localization, which is expected by us in some ultrathin films though so far there was no such report.

\begin{table}[htb]
  \centering
  \begin{minipage}[t]{\linewidth}
      \caption{The experimentally fitted prefactor $\alpha$ and phase coherence length $\mathcal{L}$ in the Hikami-Larkin-Nagaoka magnetoconductivity Eq. (\ref{HLN}). }
\label{tab:WAL}
\begin{tabular}{cccccc}
    \hline\hline
   Ref. & sample & thickness & $T$ (K)& $\alpha$  & $\mathcal{L}$ (nm) \\ \hline
 \cite{Chen10prl} &  Bi$_2$Se$_3$ & 10 nm  &  1.8 & -0.5$\sim$-0.38 & 106$\sim$237 \footnote{Calculated from $B_{\phi}\equiv\hbar/(4e\mathcal{L}^2)$.} \\ \hline
 \cite{Checkelsky11prl} &  Bi$_2$Se$_3$ & 10$\sim$20 nm &  0.3 & -0.38 &  \\ \hline
 \cite{He11prl} &     Bi$_2$Te$_3$ & 50 nm &  2 & -0.39 & 331 \\ \hline
 \cite{Liu11prb} &     Bi$_2$Se$_3$ & 2$\sim$6 QL &  1.5& -0.39 & 75$\sim$200 \\ \hline
 \cite{Wang11prb}\footnote{Electron-electron interaction was also considered in the fitting.} &     Bi$_2$Se$_3$   & 45 nm &   0.5 & -0.31  & 1100  \\
 &    (Bi,Pb)Se$_3$  &  &   & -0.35 & 640 \\ \hline
 \cite{Chen11rc} &  Bi$_2$Se$_3$ & 5$\sim$20 nm  &  0.01$\sim$2 & -1.1$\sim$-0.4 & 143$\sim\infty$  \\ \hline
  \cite{Kim11arxiv} &     Bi$_2$Se$_3$ & 3$\sim$100 QL & 1.5 & -0.63$\sim$-0.13 & 100$\sim$1000  \\\hline
  \cite{Steinberg11arxiv} &     Bi$_2$Se$_3$ & 20 nm & 0.3$\sim$100 & -1.1$\sim$-0.7 & 15$\sim$300  \\\hline
       \hline
  \end{tabular}
 \end{minipage}
\end{table}

\subsection{\label{sec:morebulk}Two surface channels and more than two bulk channels}

\begin{figure}[tbph]
\centering \includegraphics[width=0.49\textwidth]{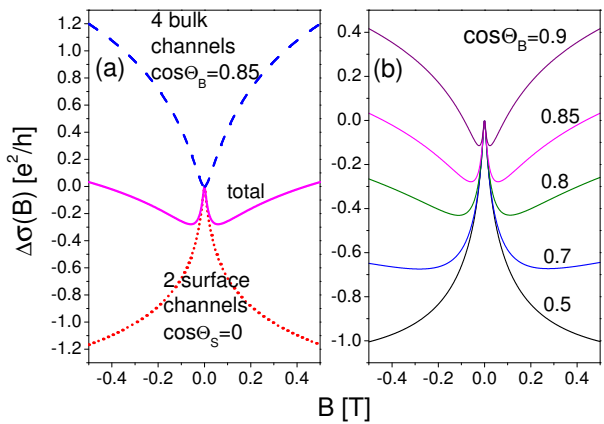}
\caption{(a) The magnetoconductivity of two surface channels with $\cos\Theta_S=0$ (dotted line) and four bulk channels with $\cos\Theta_B=0.85$ (dashed line), and their summation (solid line). (b) Total magnetoconductivity of two surface channels with $\cos\Theta_S=0$ and four bulk channels in the unitary or weak localization regime. All the other parameters are given in Fig. \ref{fig:mc_Dirac}.}
\label{fig:mc_4bulks}
\end{figure}

Now we considered the case in Fig. \ref{fig:Ek}(b), where more than two bulk channels can participate in the transport, and the surface channels have negligible gap.
Without loss of generality, we consider four bulk channels. It is fair to expect the 2D bulk subbands higher than the lowest pair to have similar descriptions as the modified Dirac model in Eq. (\ref{H3Dsub}) but with different parameters. For simplicity, we assume that the four bulk channels have the same $\cos\Theta_B$.

Because now we have unequal numbers of weak antilocalization channels from the surface bands and weak localization channels from the bulk subbands,
the weak localization behavior of the bulk channels may be hidden by the weak antilocalization of the surface channels at small magnetic field,
but the superiority in channel number of the bulk band will eventually change the trend at some finite magnetic field.
As shown in Fig. \ref{fig:mc_4bulks}, inflections in magnetoconductivity appear at finite magnetic fields, and even change the magnetoconductivity from negative to positive. These inflections will be evident signatures indicating that the bulk states could have weak localization instead of always weak antilocalization.

\section{\label{sec:summary}Summary and Discussion}

Starting with the modified Dirac model as a unified starting point, we derive the magnetoconductivity formula for both the lowest 2D bulk subbands and surface bands of a topological insulator thin film.
A crossover from the weak antilocalization to weak localization is expected, controlled by the pseudospin polarization defined by the parameters in the modified Dirac model. Unlike those in the ferromagnetic semiconductor, this crossover does not break time-reversal symmetry.
Due to their relatively large gap, we suggest that the bulk states may lie in the regimes where the unitary behavior or even the weak localization is also expected, instead of the expected weak antilocalization.
As a result, the experimentally observed weak antilocalization may be a collective result of the weak antilocalization of the surface bands and weak localization of the bulk bands. It may explain the deviation of the fitting prefactor of the Hikami-Larkin-Nagaoka formula from the expected $-1/2$ in the experiments, as well as the insensitive response of the bulk states to magnetic field or doping.
When the bulk channels outnumber the surface channels, inflection in the magnetoconductivity may appear at some finite magnetic field. The inflection will give a signature that the bulk states of a topological insulator can give weak localization, although they have strong spin-orbit coupling.

In this work, we consider only the ultrathin limit of a 3D topological insulator.
In the bulk limit, because a large number of 2D bulk subbands will contribute to the transport, the direct coupling [by the $k_z$ terms in Hamiltonian (\ref{H3D})] and scattering via impurities among them become inevitable.
In this limit, we argue that the inter-subband coupling and scattering will give extra ``Cooperon gaps" to all the vertex corrections from the maximally crossed diagrams for the 2D bulk bands.
These Cooperon gaps will reduce effectively the phase coherence length, and drive the transport via each 2D bulk subband out of the quantum diffusion regime.
As a result, we expect the 3D bulk states to have neither weak localization nor weak antilocalization, but the unitary behavior with small magnetoconductivity proportional to $B^2$.

We thank H. T. He, J. N. Wang, F. C. Zhang, Junren Shi, W. Q. Chen, and B. Zhou
for helpful discussions. This work is supported by the Research Grant
Council of Hong Kong under Grant Nos. HKU 7051/10P and HKUST3/CRF/09.


\end{document}